\documentclass[prd,twocolumn,superscriptaddress,floatfix,nofootinbib]{revtex4-1}
\usepackage[utf8]{inputenc}
\usepackage{graphicx}
\usepackage{graphics,bm}
\usepackage{amsmath}
\usepackage{amsfonts}
\usepackage{amssymb}
\usepackage{color}
\usepackage{multirow}
\usepackage{subcaption}

\newcommand{\bra}[1]{\langle #1 |}
\newcommand{\ket}[1]{\left| #1 \right\rangle}

\def\6{\langle}
\def\9{\rangle}

\newcommand\tx{\mathtt{x}}
\newcommand\ttt{\mathtt{t}}
\newcommand\tE{\mathtt{E}}
\newcommand\tp{\mathtt{p}}
\newcommand\tq{\mathtt{q}}
\newcommand\tc{\mathtt{c}}

\newcommand\tC{\mathtt{C}}
\newcommand\tL{\mathtt{L}}

\newcommand\tu{\mathtt{u}}
\newcommand\tN{\mathtt{N}}

\newcommand\tga{\mathtt{\gamma}}
\newcommand\tw{\mathtt{w}}
\newcommand\taa{\mathtt{\alpha}}

\newcommand\wtPsi{\widetilde{\mathtt{\Psi}}}

\newcommand\wtPi{\widetilde{\mathtt{\Pi}}}

\newcommand{\be}{\begin{equation}}
\newcommand{\ee}{\end{equation}}
\newcommand{\ba}{\begin{eqnarray}}
\newcommand{\ea}{\end{eqnarray}}

\usepackage{amsmath}

\begin{document}
\title{Localization of scalar quantum fields on Minkowski space-time}

\author{Vasileios I. Kiosses}
\email{kiosses.vas@gmail.com}
\affiliation{QSTAR, INO-CNR, Largo Enrico Fermi 2, I-50125 Firenze, Italy}

\begin{abstract}
A new localization scheme for Klein-Gordon particle states is introduced in the form of general space and time operators. The definition of these operators is achieved by establishing a second quantum field in the momentum space of the standard field we want to localize (here Klein-Gordon field). 
The motivation for defining a new field in momentum space is as follows. In standard field theories one can define a momentum (and energy) operator for a field excitation but not a general position (and time) operator because the field satisfies a differential equation in position space and, through its Fourier transform, an algebraic equation in momentum space. Thus, in a field theory which does the opposite, namely it satisfies a differential equation in momentum space and an algebraic equation in position space, we will be able to define a position and time operator. 
Since the new field lives in the momentum space of the Klein-Gordon field, the creation/annihilation operators of the former, which build the new space and time operators, reduce to the field operators of the latter. As a result, particle states of Klein-Gordon field are eigenstates of the new space and time operators and therefore localized on a space-time described by their spectrum. Finally, we show that this space-time is flat because it accommodates the two postulates of special relativity. Interpretation of special relativistic notions as inertial observers and proper acceleration in terms of the new field is also provided.	
\end{abstract}

\maketitle


\section{Introduction}
 
A scalar field is known to be described by the Klein-Gordon theory. Even though this description is part of the extensively verified physical theory of Standard model, it encounters certain difficulties.
Space-time localization or, in other words, the definition of general space-time operators is unquestionably one of them.
The concept of space-time localization in the context of quantum field theory has been a challenging issue since the advent of the latter.
And yet, a definite answer is still lacking \cite{N-W,Wightman,Hegerfeldt,Haag,Barat Kimball,Busch,Terno,Toller,C-K-T}.
Indeed, there is no unique way to describe localization of relativistic particles, even when the particles are unambiguously defined.
In standard formalism of quantum mechanics, time is never treated as an operator, and therefore has an entirely different description from that of space position \cite{Salecker et al,Busch et al}.
It was Pauli \cite{Pauli}, who first pointed out that a construction of a general time operator in quantum mechanics is impossible. Since time translations are generated by the Hamiltonian, any legitimate general time operator, as conjugate to the Hamiltonian, should translate energy arbitrarily.
Thus, the definition of time operator contradicts the fact that any realistic Hamiltonian has a spectrum bounded from below.
The same argument was generalised by Wightman in relativistic case (where space and time are merged into space-time) for position-time operators \cite{Wightman}.
The standard textbook answer to Pauli-Wightman objection is that space-time position in quantum field theory cannot be represented by an operator and is just a parameter, external to the theory \cite{Peres}.

Another difficulty is the infinite energy of the ground state or vacuum energy as it is known.
Even though the ground state demonstrates that the energy is bounded below, an interesting, and for the same reason disturbing, phenomenon takes place in Klein-Gordon theory (which is common in all standard quantum field theories), the ground state has infinite energy. 
The standard way to deal with this problem, is to accept the theory only up to a certain ultraviolet momentum $\tp_{max}$ and cut off the very high momentum modes.
In the absence of gravity this finite energy has no effect. The values that it can take is completely arbitrary. Traditionally, this energy is discarded by the process of ``normal-ordering''.
However, the vacuum fluctuations are very real, having measurable impact in standard physics, see Casimir effect \cite{Casimir}. 
On top of that, gravity does exist, and the actual value of the
vacuum energy has important consequences.
It turns out that the energy density of the vacuum measures the cosmological constant.
The cosmological constant problem, the discrepancy between the theoretical (from calculations of vacuum energy) and observational values of the cosmological constant, is not related to the fact that quantum theory supply a system with a huge amount of vacuum energy, since this contribution can be renormalized away, the problem is that there is no reason the resulting number to be zero \cite{Carroll}.

While, in principle, the ground state, as it is, could have solved the problem of space-time localization in quantum field theory, in the sense that the infinite vacuum energy practically renders Hamiltonian spectrum unbound from below, the fact that the current field theories cannot handle vacuum energy put forward Pauli-Wightman objection setting us incapable of defining space-time operators and therefore forcing us to relegate  position $x$ and time $\ttt$ to mere parameters.
From the discussion above, one can conclude that with regard to the aforementioned issues, the present description of Klein-Gordon field is no satisfactory and a new context able to theorize the arbitrariness of vacuum energy in terms of a new potential in order to adequately define space-time operator, while remaining consistent with the known laws of physics, would be desirable.
In this work we show that there exists a way which accomplishes this end.
The method consists of attaching to the momentum space of the free Klein-Gordon field an extra degree of freedom, interpreted as potential energy, associated it to the vacuum energy, and then introducing, in addition to the standard equation of motion of the field (i.e. Klein-Gordon equation) a second equation in the extended momentum space of the field. As we will show, the physical meaning of the new equation is the energy transfer between the free energy $E_\tp$ and the newly-established potential/vacuum energy. However, in practice, it can be interpreted as a new type of quantum fields which provides the space-time where the standard field is localized. In a different context, the same definition of the new field was used to derive Unruh effect \cite{C-K}.

More specifically we shall show that for every real field of Klein-Gordon theory, making use of its vacuum energy, it can be defined a second field in its extended momentum space (i.e. free energy and vacuum energy) which has a set of time and space operators complying with the language of second quantization. Consequently, a general multi-particle state, eigenstate of the field Hamiltonian, is also eigenstate of the new time and position operators. The corresponding space and time eigenvalues are proportional to the wave numbers of the new field and the eigenvalues of the Klein-Gordon number operator. On account of these operators, the space-time localization of the Klein-Gordon field momentum states is not, any more, external to the theory, through parameters, as standard approach supports \cite{Peres}, but on some internal degrees of freedom of the field operators.
	
From the kinematic study of the space-time localised quantum particles further insight on 
the nature of space-time and behaviour of particles states on this space-time is obtained. In particular, it is found that space-time is relativistic in nature satisfying the two postulates of special relativity. And the contribution of vacuum energy in the form of a extra potential energy renders the particle states motion accelerating.

In section \ref{Localization scheme} we introduce the new field and define the general position and time operators. We apply the new localization scheme on Klein-Gordon particle states in section \ref{Application to momentum eigenstates}. By studying the kinematics of the new space-time operators' spectrum, in section \ref{Kinematics on the space-time spectrum} we analyze the space-time where particles states are localized on. A final discussion is then presented in section \ref{In lieu of conclusions}. 
For simplicity, we consider the quantum theory of real Klein-Gordon field in two dimensions, with metric signature $(+,-)$, but our results can easily be generalised to four dimensions (see \ref{In lieu of conclusions}). Furthermore, the units are chosen such that $c = \hbar=1$, unless specified otherwise.

\section{Localization scheme}\label{Localization scheme}

A massive, scalar field $\Phi$ is known to be described by the Klein-Gordon equation  
\be
(\partial_{\ttt}^2 - \partial_{x}^2 + m^2)\Phi(\ttt,x) = 0. \label{Klein-Gordon_equation}  
\ee
$x,\ttt$ are the coordinates of (1+1) space-time where the field is defined. 
One can (second-)quantize the field by forming the field operator $\hat{\Phi}(\ttt,x)$ in the usual way and expanding in the normal mode solutions $\phi_{\tp}(\ttt,x) = e^{-i(E_{\tp} \ttt - \tp x)}/\sqrt{4 \pi E_{\tp}}$ of Klein-Gordon equation (\ref{Klein-Gordon_equation}),
\be
\hat{\Phi}(\ttt,x) = \int d\tp \left(\hat{a}_{\tp} \phi_{\tp}(\ttt,x) + \hat{a}^\dagger_{\tp} \phi_{\tp}^*(\ttt,x) \right). \label{standard_K-G}
\ee
For each value of momentum $\tp$, it corresponds the energy eigenvalue $E_{\tp}= \sqrt{\tp^2 + m^2}$.
The associated ground state $\ket{0}$ is defined by 
\be
\hat{a}_{\tp}\ket{0}=0 \qquad \bra{0}\hat{a}^{\dagger}_{\tp}=0,\quad \forall\tp
\ee
where $\hat{a}_{\tp}$ and $\hat{a}^{\dagger}_{\tp}$ are the standard annihilation and creation operators of the theory, respectively.
The Hamiltonian of the Klein-Gordon field $\Phi$ reads
\be
\hat{H}=\int d\tp \, E_\tp \,\left(\hat{a}_\tp^\dagger\, \hat{a}_\tp +\frac{1}{2} [\hat{a}_\tp, \hat{a}_\tp^\dagger]\right).\label{KG-Hamiltonian}
\ee

Let us start introducing our localization scheme by considering the momentum space of the Klein-Gordon field $\Phi$ to have as many as two independent degrees of freedom, which can characterize any point in this space.
We choose the one to be the momentum $\tp$ of the Klein-Gordon particles and the second to be some abstractly, for now, defined energy $\tE$ (remember that $c=1$ so energy and momentum share the same units). 
We work with a specific Klein-Gordon field, so we consider the field mass to have a definite value and therefore not to be one more degree of freedom in momentum space.
Technically speaking, for now, energy $\tE$ is physically meaningless.
The only known energy that the system has is the free energy $E_{\tp}$. But this energy is a function of momentum, $E_{\tp}\equiv E(\tp)$, thus the energy coordinate $\tE$ in momentum space of Klein-Gordon field cannot be that energy.
The sole purpose of $\tE$ is to unambiguously label the potential energy, related to the vacuum energy of the system. Apparently, since there is no restriction on the values that the coordinates $\tE,\tp$ can take, the domain of both $\tE$ and $\tp$ is the set of all real numbers.

Looking for a way to integrate the energy $\tE$ to the free Klein-Gordon field $\Phi$, we add an interaction between the momentum $\tp$ and $\tE$ in a form of a novel field.
Let us assume that the momentum space of the Klein-Gordon field, as described above, has its own fields in the sense that each point in momentum space is associated with a continuous field variable $\wtPsi(\tE,\tp)$, which satisfies the differential equation:
\be
\left(\hbar^2\partial_{\tE}^2-\hbar^2\partial_{\tp}^2 - \frac{1}{\kappa^2} \right)\wtPsi(\tE,\tp) = 0. \label{accelerated_diff.-equation}
\ee 
Let us explain the terms that constitute this equation. 
First we should notice that temporarily we recover the reduced Planck constant $\hbar$ to avoid confusion in our analysis below.
$\kappa\in\Re$ is just a field parameter. Later we will see that it is related to the proper acceleration of localized particles states.
The differential operators $\hbar^2\partial_{\tp}^2, \hbar^2\partial_{\tE}^2$ are the square of the linear differential operators $i\hbar{\partial}_{\tp}\equiv i\hbar\frac{\partial}{\partial \tp} $ and $-i\hbar{\partial}_{\tE} \equiv - i\hbar\frac{\partial}{\partial \tE}$, respectively.
 
To comprehends the physical meaning of these operators we need first to solve equation (\ref{accelerated_diff.-equation}). 
Mathematically speaking, eq.(\ref{accelerated_diff.-equation}) is recognized as a classical wave equation including an extra term, so its solutions are plane waves of the form
\be
\tu(\tE,\tp) = \frac{e^{i(\tE \, \widetilde{k} - \tp\, \widetilde{\omega})}}{\sqrt{4 \pi \widetilde{\omega}}} \label{planewave1}.
\ee
Factor $1/\sqrt{4\pi\widetilde{\omega}}$ was inserted for later convinience.
Plane waves $\tu$ are solutions to eq.(\ref{accelerated_diff.-equation}) in the same sense in which $\phi$ are solutions to Klein-Gordon equation (\ref{Klein-Gordon_equation}). But there is an essential difference between the two cases, which reflects the difference between the fields $\Phi$ and $\wtPsi$: while $\phi$ is defined in position space $x-\ttt$, $\tu$ is defined in momentum space $\tp - \tE$. 
Accordingly, $\ttt,x$ are the coordinate variables for $\phi$, whereas $\tp,E_{\tp}$ are the components of the wave vector with $E_{\tp}$ to obey the dispersion relation $\tE_p^2=\tp^2+m^2$. The coordinates at $\tu$, instead, are the energy $\tE$ and momentum $\tp$, while $(\widetilde{k},\widetilde{\omega})$ are the components of a wave vector with $\widetilde{\omega}$ to satisfy the dispersion relation
\be
\widetilde{\omega}^2=\widetilde{k}^2+\frac{1}{\hbar^2\kappa^2}\label{dispersion-rel1}.
\ee

Recalling the standard definition (which applies to plane waves $\phi$), a plane wave is a function of space and time coordinates, $x, \ttt$, which is proportional to 
\be
\text{plane wave} = A\, e^{i(kx-\omega \ttt)},
\ee
(where $k$ is the ordinary angular wavenumber and $\omega$ is the ordinary angular frequency). 
Apparently, according to this definition, it is rather opaque our interpretation of (\ref{planewave1}) as plane wave.
In case of matter plane wave $\phi$, this has been clarified by the de-Broglie hypothesis
\be
E_{\tp}=\hbar \omega\quad \text{and} \quad \tp=\hbar k.
\ee
However, the same hypothesis does not apply onto $\tu$, since the energy-momentum vector is already present in the argument. We need an analogous hypothesis to associate the new wave vector $(\widetilde{k},\widetilde{\omega})$ with the space-time coordinates, which is lacking. To this end acting the differential operators $i\hbar{\partial}_{\tp}$ and $-i\hbar{\partial}_{\tE}$ on the plane wave $\tu$ we get
\be
i\hbar{\partial}_{\tp} \,\tu = \hbar\widetilde{\omega}\,\tu \quad \text{and} \quad -i\hbar{\partial}_{\tE} \,\tu = \hbar\widetilde{k}\,\tu.
\ee
We postulate that the momentum derivative $i\hbar{\partial}_{\tp}$ is the position operator\footnote{In momentum representation of quantum mechanics position operator has exactly this form.}, and therefore we theorize the association of frequency $\widetilde{\omega}$ to a position $\tx_t$:
\be
\tx_t = \hbar \, \widetilde{\omega}.\label{quantum-position}
\ee
Similarly, we postulate that the energy derivative $-i\hbar{\partial}_{\tE}$ corresponds to a time operator and thus $\widetilde{k}$ is associated to time $\ttt$
\be
\ttt = \hbar \, \widetilde{k}. \label{quantum-time}
\ee
Applying eqs (\ref{quantum-position},\ref{quantum-time}) on (\ref{planewave1}) we finally get
\be
\tu(\tE,\tp) = \frac{e^{\frac{i}{\hbar}(\tE \, \ttt - \tp\, \tx_t)}}{\sqrt{4 \pi \tx_t/\hbar}} \label{planewave12}.
\ee
Evidently, if matter plane waves $\phi_\tp$ ($\hbar=1$) represent a free particle that carries momentum $\tp$ and energy $E_{\tp}$, then $\tu$ above, according to de-Broglie formulas, represents a particle with energy $-\tE$ and momentum $-\tp$.  The potential energy $\tE$ is justified because the particle is not anymore free - in section \ref{Kinematics on the space-time spectrum} we will demonstrate that the particle state has proper acceleration. The position $\tx_t$ of the particle is restricted now to the spacetime hyperboloid
\be
\tx_t^2 - \ttt^2 = \frac{1}{\kappa^2} \label{dispersion-rel2}.
\ee
Hyperboloid (\ref{dispersion-rel2}) is nothing else than the dispersion relation (\ref{dispersion-rel1}) after making use of (\ref{quantum-position},\ref{quantum-time}).

The analysis above suggests that the determination of position and time of a quantum particle turns into the eigenvalue problems:
\be
i\hbar{\partial}_{\tp} \,\tu = \tx_t\,\tu,\quad \text{and} \quad -i\hbar{\partial}_{\tE} \,\tu = \ttt\,\tu, \label{position-time_eigenequations}
\ee  
where $\tx_t$ are the position eigenvalues and $\ttt$ are the time eigenvalues.

Equations (\ref{position-time_eigenequations}) say that the position operator $i\hbar{\partial}_{\tp}$ is the generator of momentum change and the time operator $-i\hbar{\partial}_{\tE}$ is the generator of (potential) energy change. More specifically, if we consider the plane wave $\tu(\tE, \tp+\tq)$ and expand it into power series, we obtain
\ba
\tu(\tE, \tp+\tq) &=& \tu(\tE, \tp) + \tq \partial_{\tp}\tu(\tE, \tp)+\frac{1}{2}\left(\tq \partial_{\tp}\right)^2\tu(\tE, \tp)+\ldots \nonumber\\
&=& e^{-\frac{i\tq}{\hbar}\cdot i\hbar{\partial}_{\tp}} \,\tu(\tE, \tp)\nonumber \\
&=& e^{-\frac{i}{\hbar}\,\tq\cdot \tx_t} \,\tu(\tE, \tp).\label{unitay-momentum}
\ea
Similarly, the plane wave $\tu(\tE-e, \tp)$ expanded into power series gives
\ba
\tu(\tE-e, \tp) &=& \tu(\tE, \tp) -e \partial_{\tE}\tu(\tE, \tp)+\frac{1}{2}\left(-e \partial_{\tE}\right)^2\tu(\tE, \tp)+\ldots \nonumber\\
&=& e^{-\frac{i e}{\hbar}\cdot (-i\hbar{\partial}_{\tE})} \,\tu(\tE, \tp)\nonumber \\
&=& e^{-\frac{i}{\hbar}\,e\cdot \ttt} \,\tu(\tE, \tp).\label{unitay-energy}
\ea
The unitary operators $U_{\tx_t}(\tq):=e^{-\frac{i\tq}{\hbar}\cdot i\hbar{\partial}_{\tp}}$ and $U_{\ttt}(e):=e^{-\frac{i e}{\hbar}\cdot (-i\hbar{\partial}_{\tE})}$ built by the position and time operators, respectively, responsible for the transformation of $\tu(\tE,\tp)$, do not commute since position $\tx_t$ and time $\ttt$, are not independent degrees of freedom, see eq.(\ref{dispersion-rel2}). Their non-commutativity will be demonstrated later in terms of creation and annihilation operators.
For now we present the relation
\be
\tu(\tE,\tp\pm\tq) = \tu(\tE-f(\tq,\kappa),\tp), \label{unitary_momentum-energy}
\ee
which is easily derived combining eqs.(\ref{unitay-momentum},\ref{unitay-energy}) with (\ref{dispersion-rel2}), assuming the function
\be
f(\tq,\kappa) = \frac{\pm\tq}{\ttt}\sqrt{\ttt^2 + \frac{1}{\kappa^2}}.\label{fqk}
\ee
The double sign in front of $\tq$ covers the fact that solving (\ref{dispersion-rel2}) for $\tx_t$ we get $\tx_t=\pm \sqrt{\ttt^2 + 1/\kappa^2}$.
Remember that we have justified the introduction of the new field equation ($\ref{accelerated_diff.-equation}$), arguing that in this way the actual momentum $\tp$ of Klein-Gordon particles is integrated with the conjectural potential energy $\tE$. 
Eq.(\ref{unitary_momentum-energy}) explicitly verifies that this is the case.
The plane wave of a particle with potential energy $\tE$ and momentum $\tp$ increased by $\tq$ is the same with the plane wave of a particle with the initial momentum $\tp$ and potential energy decreased by a function of $\tq$ and parameter $\kappa$.
So, one can infer that the physical meaning of $\wtPsi$ is the energy transfer between potential energy $\tE$ and free energy $E_{\tp}$ (``kinetical'' in the sense that depends on momentum, $E_{\tp}=E_{\tp}(\tp)$).

It is important to underline the fact that we have identified the eigenvalues of time operator $-i\hbar\partial_{\tE}$ as the time parameter which appears in standard Klein-Gordon field equation, by using the same letter, $\ttt$, in both cases. Doing this we assert that the time that appears in Klein-Gordon equation (\ref{Klein-Gordon_equation}) is not anymore a parameter but an operator.
We cannot argue the same regarding the position parameter $x$ and position eigenvalue $\tx_t$ for two reasons, both rooted in relation (\ref{dispersion-rel2}). Whereas $x$ and $\ttt$ are independent degrees of freedom, $\tx_t$ and $\ttt$ are not. Secondly, in contrast to $x$ which can take any value from real line, $\tx_t$ is restricted to the spacetime hyperboloid (\ref{dispersion-rel2}).

From now on, natural units comprising $\hbar=1$ (in addition to $c=1$) are used allowing time $\ttt$, wavenumber $\widetilde{k}$, length $\tx_t$ and angular frequency $\widetilde{\omega}$ to be used interchangeably.
Thus, henceforward we recognize as dispersion relation the equation (\ref{dispersion-rel2}), and as plane wave the relation
\be
\tu_{\ttt}(\tE,\tp) = \frac{e^{i(\ttt \,\tE - \tx_t\, \tp)}}{\sqrt{4 \pi \tx_t}} \label{planewave2}.
\ee

To write down the most general solution of (\ref{accelerated_diff.-equation}), we need to construct a complete, orthonormal set of plane waves (modes) in terms of which any solution may be expressed. But to make sense of orthonormal we have to define an inner product on the space of solutions to the equation (\ref{accelerated_diff.-equation}). The appropriate inner product is expressed as an integral over a constant-momentum curve ${C}$,
\be
(\tu,\mathtt{v}) := i \int_{{C}}    
d{\tE} \left(\tu^* \, \partial_{\tp} \mathtt{v}
-\mathtt{v} \, \partial_{\tp}\tu^*\right). \label{eq.inner-prod}
\ee
The functions $\tu$ and $\mathtt{v}$
are plane waves of the form of Eq.(\ref{planewave2}). 
It is easily verified that the plane waves set up an orthonormal set under this product,
\ba
(\tu_{\ttt'},\tu_{\ttt}) &=&
\delta(\ttt - \ttt') \nonumber \\
(\tu^*_{\ttt'},\tu^*_{\ttt}) &=& 
-\delta(\ttt - \ttt') \label{eq.inner-prod2}\\
(\tu_{\ttt'},\tu^*_{\ttt}) &=&0.\nonumber
\ea
Note that the inner product (\ref{eq.inner-prod}) is not positive definite. From (\ref{eq.inner-prod2}) we can choose the positive-frequency mode as $\tu_{\ttt}$ and, consequently, the negative-frequency one as $\tu^*_{\ttt}$.
Based on this inner product, $\wtPsi$ can be expressed as a Fourier expansion of these normal modes
\be
\wtPsi(\tE,\tp) = \int d\ttt \left(\tc_\ttt \, \tu_{\ttt}(\tE,\tp) + \tc^\dagger_\ttt \, \tu_{\ttt}^*(\tE,\tp)\right).
\label{Field_expansion-Fourier-K_G-Acc1}
\ee
with the Fourier coefficients to be defined by 
\be
\tc_\ttt = (\tu_{\ttt},\wtPsi) \qquad \text{and} \qquad \tc^\dagger_\ttt = -(\tu^*_{\ttt},\wtPsi) \label{creat-annih1}.
\ee

The field $\wtPsi$ can be quantized according to the rules of second quantization by promoting $\tc_\ttt, \tc_\ttt^\dagger$ to annihilation and creation operators, respectively, and satisfying the usual commutation relation for the raising and lowering operators,
\be
[\hat{\tc}_\ttt, \hat{\tc}_{\ttt'}^\dagger]=\delta(\ttt-\ttt'). \label{commutation -time}
\ee
Due to Eq.(\ref{Field_expansion-Fourier-K_G-Acc1}), the commutation relations (\ref{commutation -time}) in coordinate space are equivalent to the commutation relations
\be
\left[\hat{\wtPsi}_{\tp}(\tE),\partial_{\tp} \hat{\wtPsi}_{\tp}(\tE')\right] = i \delta(\tE - \tE'),\label{comm.-rel-momentum}
\ee  
in momentum space. Here, index $\tp$ stands for a curve in momentum space of constant $\tp$. This relation implies that operators at equal momentum commute everywhere except at coincident potential-energy points.

Creation operators $\hat{\tc}_\ttt^\dagger$ and annihilation operators $\hat{\tc}_\ttt$ following the general machinery of second quantization should be operator-valued distributions. Furthermore, since $\hat{\wtPsi}(\tE,\tp)$ lives in momentum space of the Klein-Gordon theory, its Fourier coefficients, the quantities $\hat{\tc}_\ttt, \hat{\tc}_\ttt^\dagger$, should live in coordinate space $\ttt-\tx_t$, or the one-dimensional space $\ttt$ due to (\ref{dispersion-rel2}). 
Remember that we have identified this $\ttt$ with the time parameter appearing in the field operator $\hat{\Phi}(\ttt,x)$, which, again according to second quantization, is an operator-valued distribution.
Thus, we infer that the time annihilation and creation operators can be represented by the projections of Klein-Gordon field operator, $\hat{\Phi}(\ttt,x)$ and $\hat{\Phi}^\dagger(\ttt,x)$ respectively, for fixed value of position:
\be
\hat{\Phi}_{x}(\ttt) :=\left.\hat{\Phi}(\ttt,x_0)\right|_{x_0 \rightarrow x} =\hat{\tc}_\ttt \label{Phi_to_tc}
\ee
and
\be
\hat{\Phi}_{x}^\dagger(\ttt) :=\left.\hat{\Phi}^\dagger(\ttt,x_0)\right|_{x_0 \rightarrow x} =\hat{\tc}^\dagger_\ttt. \label{Phid_to_tcd}
\ee
Apparently, since $x$ does not appear on the solutions to (\ref{accelerated_diff.-equation}), we assert that (\ref{Phi_to_tc},\ref{Phid_to_tcd}) hold for every $x$.
The index notation is used to distinguish the parameter $x$ from the real variable $\ttt$.

Another argument that supports the connection between the time creation/annihilation operators and Klein-Gordon field operator is the following.
Using equation (\ref{accelerated_diff.-equation}) for $\hat{\wtPsi}$ and the fact that $\tu_{\ttt}$ is solution to the same differential equation one immediately verifies, through an integration by parts, that $\hat{\Phi}_{x}(\ttt)$ and $\hat{\Phi}_{x}^\dagger(\ttt)$ are constant in momentum $\tp$. This is consistent with the fact that the Klein-Gordon field operator in standard framework is independent of momentum since according to (\ref{standard_K-G}) $\hat{\Phi}(\ttt,x)$ is in superposition of all the possibles momentum eigenstates.

Operators $\hat{\Phi}(\ttt,x)$ and $\hat{\Phi}^\dagger(\ttt,x)$ acquire a dual role. 
As Klein-Gordon field operators, from expansion (\ref{standard_K-G}), they create Klein-Gordon particles as superposition of momentum eigenstates, at specific space-time position $(\ttt,x)$.
At the same time, as Fourier coefficients in the expansion of the wave operator $\hat{\wtPsi}(\tE,\tp)$, updated to
\be
\hat{\wtPsi}(\tE,\tp) = \int d\ttt \left( \hat{\Phi}_{x}(\ttt) \, \tu_{\ttt}(\tE,\tp) + \hat{\Phi}_{x}^\dagger(\ttt) \, \tu_{\ttt}^*(\tE,\tp)\right),
\label{Field_expansion-Fourier-K_G-Acc}
\ee
they have the function to create and annihilate time eigenstates $\tu_{\ttt}$ and $\tu^*_{\ttt}$.
Mathematically, this dual role is translated into two commutation relations that $\hat{\Phi}$ should satisfy. With respect to the first case, the equal-time standard canonical commutation relation 
\be
\left[\hat{\Phi}_{\ttt}(x),\partial_\ttt\hat{\Phi}_{\ttt}(x')\right] =i \delta(x-x'),\label{cons-tim_com_rel}
\ee 
and in connection with the second case, the commutation relation
\be
\left[\hat{\Phi}_{x}(\ttt),\hat{\Phi}^\dagger_{x}(\ttt') \right] = \delta(\ttt-\ttt').   \label{cons-pos_com_rel}
\ee
which is typical for any set of creation and annihilation operators.

A key physical question is what are the observable quantities defined by this new field theory? 
In standard matter fields (like Klein-Gordon, Dirac), the simplest and most important such object is the overall Hamiltonian, which represents the total energy of the system. 
Passing to the new field $\wtPsi$, the same formalism can be maintained
by considering the total ``Hamiltonian''
\be
\hat{X} = \int d\ttt \,  \tx_t\, \hat{\Phi}_{x}^\dagger(\ttt)\,\, \hat{\Phi}_{x}(\ttt), \label{X-position}
\ee
which has dimensions of length, $\tx_t = \sqrt{\ttt^2 + 1/\kappa^2}$.
In addition to the Hamiltonian, from our theory it follows also the operator
\be
\hat{T} = \int d\ttt\,  \ttt\,\,\hat{\Phi}_{x}^\dagger(\ttt)\,\, \hat{\Phi}_{x}(\ttt) . \label{T-time}
\ee 
which has dimensions of time, see also \cite{C-K2}. For more details on the derivation see Appendix.

\section{Localization of momentum states} \label{Application to momentum eigenstates}

We define a $\tN-$particle state vector by
\be
\ket{\tN}:= \left(\hat{\Phi}^\dagger(\ttt,x)\right)^{\tN}\ket{0}, \label{multi-particle states}
\ee
or, once we apply the expansion (\ref{standard_K-G}), 
\be
\ket{\tN} =  \int d\tp_1\ldots d\tp_\tN \, \phi_{\tp_1}^*(\ttt,x)\ldots \phi_{\tp_\tN}^*(\ttt,x) \, \ket{\tp_1,\ldots,\tp_\tN}, \label{superposition_momentum_eigenstates}
\ee
where $\ket{\tp_1,\ldots,\tp_\tN}:=\hat{a}_{\tp_1}^\dagger\ldots \hat{a}_{\tp_\tN}^\dagger\ket{0}$.
This state is a superposition of $\tN$-particle momentum eigenstates and according to the standard Klein-Gordon theory describes a system of $\tN$ particles, that at time $\ttt$ are localized in coordinate space at point $x$.

Since in standard Klein-Gordon theory there is not exist a legitimate position operator the interpretation above can be confirmed if we introduce the local particle-number operator \cite{Greiner-R}
\be
\hat{\tN}_V(\ttt):=\int_V dx\, \hat{\Phi}^\dagger(x,\ttt)\,\, \hat{\Phi}(x,\ttt).
\ee
This operator, in contrast to standard number-operator $\hat{\tN}=\int d\tp \,\hat{a}_\tp^\dagger\, \hat{a}_\tp$, will, in general, depend on time due to the possible spreading of the field. The index $V$ on the integral indicates the volume limit that the integration extends over (which of course can be infinitesimally small).
Because of commutation relations (\ref{cons-tim_com_rel}) the operator $\hat{\tN}_V$ satisfies the relation
\be
\left[\hat{\tN}_V, \hat{\Phi}^\dagger(x,\ttt)\right] =\left\lbrace 
         \begin{aligned}
         	&\hat{\Phi}^\dagger(x,\ttt)&\quad x\in V \\
         	&0& \quad x\notin V
	     \end{aligned}
	\right.
\ee
From (\ref{multi-particle states}) and the fact that $\hat{\tN}_V\ket{0}=0$, it follows that $\ket{\tN}$ is an eigenstate of the number-operator $\hat{\tN}_V$ with two possible eigenvalues, $1$ if $x$ is contained in $V$, and $0$ if it does not.

This is the picture we have regarding the localization of $\ket{\tN}$ based on standard field theory (of the same nature result we obtain working with position POVMs, see for example \cite{C-K-T}).
However the notion of localized particle states is considerably improved applying the localization scheme, introduced at the previous section.
Considering the action of $\hat{T}\equiv \hat{T}^0$ and $\hat{X}\equiv \hat{T}^1$ (named together as $\hat{T}^\mu$, $\mu=0,1$) on the state $\ket{\tN}$, we find
\be
\hat{T}^\mu\ket{\tN} = t^\mu\tN \ket{\tN}, \label{space-time_eigenequation}
\ee
where $\ttt^\mu:=(\ttt,\tx_t)$.
So we verify that the multi-particle state $\ket{\tN}$ given by (\ref{multi-particle states}) is an eigenstate of space-time operator $\hat{T}^\mu$ with eigenvalues $t^\mu\tN$. The components of the space-time vector $t^\mu$ are comprising by the time parameter $\ttt$ appearing in field operator $\hat{\Phi}^\dagger(\ttt,x)$, and the position $\tx_t=\sqrt{\ttt^2+1/\kappa^2}$ which corresponds to this time. $\tN$ is the eigenvalue of the number operator $\hat{\tN}$ in momentum space, or equivalently to number operator 
\be
\hat{\tN}_x =\int d\ttt \, \,\hat{\Phi}_{x}^\dagger(\ttt)\,\, \hat{\Phi}_{x}(\ttt) \label{number_operator-coor_sp}
\ee
in coordinate space, which is easily concluded from equations (\ref{X-position}) and (\ref{T-time}).
Note that contrary to operator $\hat{\tN}_V$, $\hat{\tN}_x$ does not depend on time, since the integration is performed over time, but on position $x$.

Identifying the space-time localization of particle state $\ket{\tN}$ with the product $t^\mu \tN$ leads to a remarkable conclusion. 
If a single particle state, $\ket{1}$, is localized in space-time position $(\ttt,\tx_t)$, then a two-particle state, $\ket{2}$, is localized in position $(2\,\ttt,2\,\tx_t)$, and a $\tN$-particle state, $\ket{\tN}$, is localized in spacetime position $(\tN\,\ttt,\tN\,\tx_t)$.
One can gain a more clear picture of this calculating the commutators between the space-time operators $\hat{T}^\mu$ and the field operators $\hat{\Phi}$, $\hat{\Phi}^\dagger$:
\be
\begin{aligned}
	\left[\hat{T}^\mu,\hat{\Phi}_{x}^\dagger(\ttt)\right]&=\ttt^\mu \hat{\Phi}_{x}^\dagger(\ttt) \\
	\left[\hat{T}^\mu,\hat{\Phi}_{x}(\ttt)\right]&= -\ttt^\mu \hat{\Phi}_{x}(\ttt).
\end{aligned}
\ee
These relations tell us that there are the field operators $\hat{\Phi}$ and $\hat{\Phi}^\dagger$ which transfer us between space-time eigenstates.
Starting from the eigenstate $\ket{\tN}$ located at $t^\mu\tN$, according to equation (\ref{space-time_eigenequation}), we can construct all the others localized eigenstates by acting with $\hat{\Phi}$ and $\hat{\Phi}^\dagger$,
\be
\begin{aligned}
\hat{T}^\mu \, \hat{\Phi}_{x}^\dagger(\ttt) \ket{\tN}&=\left(t^\mu+1\right)\tN\ket{\tN} \\
\hat{T}^\mu \, \hat{\Phi}_{x}(\ttt) \ket{\tN}&=\left(t^\mu-1\right)\tN\ket{\tN}.
\end{aligned}
\ee
So, the field system we study has not only a ladder of energy states (produced by the action of creation and annihilation operators $\hat{a}_\tp^\dagger, \hat{a}_\tp$ on the ground state $\ket{0}$) but also a ladder of space-time states constructed by the action of field operators $\hat{\Phi}^\dagger,\hat{\Phi}$ on the same ground state $\ket{0}$.
In other words, we found that $\hat{\Phi}_{x}^\dagger(\ttt)$ ($\hat{\Phi}_{x}(\ttt)$) acting on a particle state adds (subtracts) a space-time interval $t^\mu$ by adding (subtracting) a particle.

Note that quantum states, which correspond to quantum particles of different number, cannot be ascribed to the same space-time position.
This conclusion is mathematical ensured by the fact that the number operator $\hat{\tN}_x$ commutes with the space-time operator $\hat{T}^\mu$, 
\be
[\hat{T}^\mu,\hat{\tN}_x] = 0.
\ee

We end this section with two important comments.
Space-time operators depend on the the position $x$, $\hat{T}^{\mu}_{x}$ (we have already mentioned that this is the case for number operator $\hat{\tN}_x$), since according to their definition (\ref{X-position},\ref{T-time}), we integrate the field operators over time $t$ but not over space $x$. This means that our localization scheme does not localize a particle state on the $x$ dimension. Therefore $x$ remains a parameter, external to our theory. In what follows we omit the subscript $x$ to reduce the clutter. However, as we will show in the next section, space-time constructed merely by the spectrum $t^\mu\tN$ is physical enough since, by definition, is relativistic, in the sense that its structure satisfies both postulates of special relativity.

Due to the nature of quantum mechanics we can also get linear combinations of particles states with different number of particles
\be
\ket{\tL}=\sum_{i} \, \tC_i \ket{\tN_i}.
\ee 
where $\sum_i |\tC_i|^2=1$.
$\ket{\tN_i}$ is the eigenstate of the total number operator $\hat{\tN}$ (or $\tN_x$), which describes a field theory with $\tN_i$ particles.
Thus, in case of superposition of quantum field theories it holds
\ba
\langle \hat{\tN}\rangle_\tL&:=&\bra\tL \hat{\tN} \ket{\tL} = \sum_i p_i \tN_i \\ 
\langle \hat{T}^\mu\rangle_\tL &:=&\bra\tL \hat{T}^\mu \ket{\tL}= \sum_i p_i \, (t^\mu \tN_i)
\ea
where $p_i=|\tC_i|^2$.
The probabilities appeared in the expectation values represent some classical uncertainty about the particle state and consequently about its space-time location.
This uncertainty arises because the calculated expectation values are for a superposed state and therefore, should disappear in case of distinct states. Indeed
\be
\langle \hat{\tN} \rangle_\tN  = \tN \quad \text{and} \quad \langle \hat{T}^\mu \rangle_\tN = t^\mu \tN.
\ee

\section{Kinematics on the space-time spectrum}\label{Kinematics on the space-time spectrum}

In this section we investigate the kinematics on space-time $t^\mu\tN$ based merely on the field properties of space-time operators $\hat{T}^\mu$, as described in section \ref{Localization scheme}.
We will see that this kinematics accommodates both of special relativity postulates: relativity principle and a universal speed limit.
More specifically, after we interpret the relativistic notions of inertial observers (or frames of references) and relative velocities in field terms, we derive the derive Lorentz transformations as a direct consequence of our framework.
Using the machinery of special relativity then, we proceed further and provide an explicit relation between the field parameter $\kappa$ and proper acceleration.

Let us consider a slightly more complex case, taking two copies of the same Klein-Gordon field operator at two different times: $\hat{\Phi}_x^\dagger(\ttt)$ and $\hat{\Phi}_x^\dagger(\ttt')$ (since position $x$ does not affect our new localization procedure, we take it the same in both cases).
In this example, the two quantum states we consider are $\ket{\tN}_{\ttt}:= \hat{\Phi}_x^\dagger(\ttt)^{\tN}\ket{0}$ and $\ket{\tN}_{\ttt'}:= \hat{\Phi}_x^\dagger(\ttt')^{\tN}\ket{0}$, which according to our interpretation, particle state $\ket{\tN}_{\ttt}$ is localized at space-time position 
\be
t^\mu\tN\equiv(X_\tN,T_\tN), \label{space-time position1}
\ee
while particle state $\ket{\tN}_{\ttt'}$ is localized at position 
\be
t'^\mu\tN\equiv(X'_\tN,T'_\tN).\label{space-time position2}
\ee
The fact that both quantum states describe the same number particles at different times is consistent with our result at section \ref{Application to momentum eigenstates} that the number operator $\hat{\tN}_x$ does not depend on time.
Since the field operator $\Phi^\dagger_x$ is the same in both situations, we claim that the two space-time positions are the space-time coordinates of the same event (i.e. particle state) as measured by two observers in relative motion, each observer using its own coordinate system.

Suppose the observer $\mathcal{O}$ uses the coordinates $(X_\tN,T_\tN)$ and that another observer $\mathcal{O}'$ with coordinates $(X'_\tN,T'_\tN)$ is moving with velocity $\tw$ in the positive direction as viewed from $\mathcal{O}$.
Evidently, both coordinate systems are quantum observables described by the space-time operators $\hat{T}^\mu$.
Since any observer is a coordinate system for space-time, and since all observers measure the same space-time events, it should be possible to draw the coordinate lines of one observer on the coordinate lines of another observer. So, observer $\mathcal{O}'$ will be a point in the coordinate lines of $\mathcal{O}$. This point in a time interval $dT_\tN$ changes its position by $dX_\tN=\tw dT_\tN$.

The physical meaning of the relative velocity $\tw$ between the two observers in the field context of section \ref{Localization scheme} can be conceived recalling that the quantities $\ttt,\tx_t$ originally were the wave numbers of field $\wtPsi$. 
In the language of wave mechanics we know that the derivative of (angular) frequency (i.e. $\tx_t$) with respect to the (angular) wavenumber (i.e. $\ttt$) is equal to the group velocity of the wave.
In this regard, the relative velocity $\tw$ between the two observers is nothing else than the group velocity of the field $\wtPsi$.
For definite $\tN$, as the case we consider here, $\tw$ reduces to
\be
\tw:=\frac{dX_\tN}{dT_\tN}=\frac{d\tx_t}{d\ttt}.
\ee

Let us then calculate the ``Lorentz distance'' square of two space-time points: the particle state location and the origin in each coordinate system. Making use of the dispersion relation (\ref{dispersion-rel2}), which of course is satisfied by both coordinate systems $t^\mu$ and $t'^\mu$, we get
\be
\begin{aligned}
	\mathcal{O}:\quad S^2&:=X^2_\tN-T^2_\tN=\frac{\tN^2}{\kappa^2},\\
	\mathcal{O}':\quad S'^2&:=X'^2_\tN-T'^2_\tN=\frac{\tN^2}{\kappa^2}.
\end{aligned}\label{spacetime_interval}
\ee
This result is really interesting. It demonstrates that, although the position and time of the event differ for measurements made by different observers, the spacetime interval of the event from the origin in each observer's coordinate system is the same, provided of course that the number of particles are the same in each frame, which is the only case we examine here. 
Hence, a new reading of eqs.(\ref{spacetime_interval}) should be the following
\be
S^2 = S'^2,\qquad \forall\kappa,\tN \label{C6}
\ee
which implies the identity
\be
X^2_\tN -T^2_\tN = X'^2_\tN -T'^2_\tN.\label{C7}
\ee
In classical mechanics identities of the form $\ref{C7}$ are of fundamental importance since they summarize the two postulates of special relativity \cite{Schutz}. 
In spite of the traditional and conceptually very convenient use of light signals in the derivation of these identities, in our case the derivation is quite independent of the existence of light signals, or actually of any real-world effect that travels at the speed of light.
We could say that it stems from the fact that all spacetime field pairs $(\ttt,\tx_t),(\ttt',\tx_t'),\cdots$ are emanated from a single field $\wtPsi$, or more practically expressed, share the same field parameter $\kappa$. This statement becomes more clear rewriting the dispersion relation (\ref{dispersion-rel2}) we derived from studied $\wtPsi$ as
\be
\tx_t^2 - \ttt^2 = 1/\kappa^2, \quad \forall(\ttt,\tx_t).
\ee

It is straightforward from the identity (\ref{C7}) to show that the space-time coordinates $T_\tN,X_\tN$ are related to the space-time coordinates $T'_\tN,X'_\tN$ by the transformations
\be
X'_\tN = \tga (X_\tN - \tw\, T_\tN),\qquad T'_\tN = \tga (T_\tN - \tw\, X_\tN)\label{C-LT}
\ee
where $\tw$ is the relative velocity between the two observers, $\mathcal{O}$ and $\mathcal{O}'$. To construct transformations (\ref{C-LT}), we have introduced the quantity $\tga = (1-\tw^2)^{-1/2}$. 
If we consider two spacetime events, $\mathcal{A}$ and $\mathcal{B}$, then $\Delta X_\tN$, $\Delta T_\tN$ denote the finite coordinate differences $X^\mathcal{A}_\tN-X^\mathcal{B}_\tN$, $T^\mathcal{A}_\tN-T^\mathcal{B}_\tN$. In that case, by successively replacing the coordinates $\mathcal{A}$ and $\mathcal{B}$ into (\ref{C-LT}) and subtracting, we get the transformation
\be \label{C-LT3}
\begin{aligned}
	\Delta X'_\tN &= \tga (\Delta X_\tN - \tw \Delta T_\tN),\\ 
	\Delta T'_\tN &= \tga (\Delta T_\tN - \tw \Delta X_\tN). 
\end{aligned}
\ee
If, in place of differences, we are forming differentials, we obtain identical transformations with the above but in the differentials:
\be \label{C-LT4}
\begin{aligned}
	dX'_\tN &= \tga (dX_\tN - \tw dT_\tN),\\ 
	dT'_\tN &= \tga (dT_\tN - \tw dX_\tN). 
\end{aligned}
\ee
Evidently, (\ref{C-LT}-\ref{C-LT4}) have the form of classical Lorentz transformations.

The derivation of these Lorentz transformations are conceptually very different from the derivation of the similarly named transformations in classical physics. There, they are the logical consequence of the two Einstein's postulates, here it is a consequence of the field $\wtPsi$.

Let us consider once again the two observers $\mathcal{O}$ and $\mathcal{O}'$, but this time we put the particle state $\ket{\tN}$ to move non uniformly relative to both frames. The path that the state follows can be considered as a succession of the aforementioned space-time events.  
Position and time differentials allow us to calculate the velocities $v$ and $v'$, and the accelerations $g$ and $g'$, of the state in $\mathcal{O}$ and $\mathcal{O}'$, respectively. They are simply defined as
\ba
\mathcal{O}: \,& v:=\frac{dX_\tN}{dT_\tN} \quad \text{and}\quad g:=\frac{d^2 X_\tN}{dT^2_\tN} \label{C10}\\
\mathcal{O}':& \,\,v':=\frac{dX'_\tN}{dT'_\tN} \quad \text{and}\quad g':=\frac{d^2 X'_\tN}{dT'^2_\tN}.\label{C11}
\ea  
Substituting from (\ref{C-LT4}) into (\ref{C11}), and considering, in particular, the rectilinear motion, which in one dimension reads $\tw=v$, yields the acceleration transformation formula:
\be
\taa= \tga^3\,g. \label{C12}
\ee
where we have defined the proper acceleration $\taa$ of $\ket{\tN}$ as that which is measured in its rest-reference frame (in our case $\taa=g'$). 
Noticing that the right-hand side of (\ref{C12}) is equivalent to $d(\tga\tw)/dT_\tN$ and integrating twice, setting the constant of integration equal to zero in both cases, yields the following equation
\be
X^2_\tN-T^2_\tN = \frac{1}{\taa^2}. \label{CC13}
\ee
This equation represents an hyperbolic path in the coordinate system of observer $\mathcal{O}$. It describes the uniformly accelerated motion of $\ket{\tN}$, and it is of particular importance because when one analyzes the hyperbolic path from the perspective of transformations (\ref{C-LT}), due to the invariance (\ref{C7}), eq.(\ref{CC13}) translates into itself in any reference frame $\mathcal{O}'$, that is $X'^2_\tN-T'^2_\tN = \frac{1}{\taa^2}$.

From eq.(\ref{CC13}), using the definition (\ref{space-time position1}) we get
\be
\tx_t^2-\ttt^2 = \frac{1}{\taa^2\tN^2}. \label{C13}
\ee
Comparing (\ref{C13}) with dispersion relation (\ref{dispersion-rel2}) we find that the field parameter $\kappa$ is proportional to the proper acceleration $\taa$ of the quantum state $\ket{\tN}$ and the eigenvalue of the number operator $\hat{\tN}$: 
\be
\kappa=\pm \taa \, \tN.\label{CC14}
\ee

\section{In lieu of conclusions} \label{In lieu of conclusions}
In this work we have proposed a new localization scheme for scalar fields (chosen as the simplest among quantum field theories for our arguments to be more clear) in an attempt to tackle a structural problem of current theory, the lack of general space-time operators.
The idea behind our suggested localization mechanism can be summarized as follows:
A standard, free, quantum field is defined by a differential equation in position space, or equivalently, (through Fourier transforms) by an algebraic equation in momentum space. Therefore, the resulted quantum particles are described by momentum eigenstates with well-defined momentum and energy but not well-defined position and time. 
Thus, let us add a second, new field with exactly the opposite properties, namely defined by a differential equation in momentum space and its quantum excitations to be represented by position eigenstates with well-defined posiiton and time. 

In general there is no any correlation between the two fields, unless one considers the new field in the momentum space of the standard field (or alternatively, the standard field in the posiiton space of the new field). This is the case we have considered here and found that doing so the annihilation (creation) operator of the new field coincide with the Klein-Gordon field operator (its conjugate transpose) and, as a consequence, the position eigenstates of the new field are actually superposed momentum eigenstates of the Klein-Gordon field. In other words, linear superposition of momentum states are localized in space-time since they are also space-time eigenstates.

It turns out that the new field $\wtPsi$, defined in the extended momentum space of Klein-Gordon field, is intimately related to the manifestation of space-time on which standard particle states live.
In that case, the properties of the derived space-time should be determined from the form of differential equation that defines $\wtPsi$. As we showed in this work, to derive a Minkowski space-time, i.e. space-time that satisfies the two postulates of special relativity, the field equation should be of the form (\ref{accelerated_diff.-equation}). It will be interesting to extend our analysis and investigate under which circumstances particle states can be localized in curved space-time.

In this work, we have considered the field $\wtPsi$ in $1+1$ momentum space, $(\tE,\tp)$ and, as consequence, an $1+1$-dimensional quantum space-time, $(T_\tN,X_\tN)$, emerged.
However, our argument can be easily generalised to physical dimensions: $(T_\tN:=\tN\ttt,X_\tN:=\tN\tx_t^1,Y_\tN:=\tN\tx_t^2,Z_\tN:=\tN\tx_t^3)$ respectively. 
All is needed is to extend differential equation (\ref{accelerated_diff.-equation}) to 
\be
\left(\hbar^2\partial_{\tE}^2-\hbar^2\partial_{\vec{\tp}}^2 - \frac{1}{\kappa^2} \right)\wtPsi(\tE,\vec{\tp}) = 0,
\ee 
with normal mode solutions modified to
\be
\tu_{\ttt}(\tE,\vec{\tp}) = \frac{e^{i(\ttt \,\tE - \vec{\tx_t}\, \vec{\tp})}}{\sqrt{4 \pi |\vec{\tx_t}|}} .
\ee
Where $\vec{\tp}=(\tp_1,\tp_2,\tp_3)$ and $\vec{\tx_t}=(\tx_t^1,\tx_t^2,\tx_t^3)$.
From this, every part of our analysis can be obtained.

The problem we have addressed in this work is not confined only to Klein-Gordon fields. Every field theory shares the same problem. In this sense, our approach presented here, which aims to deal with the difficulties in localization of Klein-Gordon particle states, can be equally well applied to the rest of quantum field theories. The implementation of our approach to Dirac fields is left for a later communication.

\noindent
\section*{Acknowledgments} 
I acknowledge financial support from Instituto Nazionale di Ottica - Consiglio Nazionale delle Ricerche (CNR-INO).

\appendix

\section{Hamiltonian description in generalized momentum space}
The differential equation (\ref{accelerated_diff.-equation}) is a wave equation with an extra term. In 2-dimensional spaces with coordinates $(x,t)$, we know that each solution of the d'Alembert wave equation $(\partial_t^2-\partial_x^2)F(t,x) = 0$ is a function in spatial part, $x$, and $C^\infty$-dependent on the temporal part, $t$, as a parameter. Of course, $F(t,x)$ can also be regarded as a function in $t$ that is $C^\infty$-dependent on $x$ as a parameter. While the set of solutions to the Klein-Gordon equation (\ref{Klein-Gordon_equation}) are parameterized according to the first case, this form of parameterization for the function $\wtPsi(\tE,\tp)$ leads the obtained frequency to acquire an imaginary part, resulting in an unstable (exponential) growth. Thus, we choose to associate with $\wtPsi(\tE,\tp)$ the family of functions
\be
\wtPsi_{\tp}(\tE) = \left.\wtPsi(\tE,\tp')\right|_{\tp'=\tp},
\ee
which are $C^\infty$-dependent on $\tp$ and satisfy the differential equation
\be
\partial_\tp^2\,\wtPsi_{\tp}(\tE) = \left(\partial_\tE^2- \frac{1}{\kappa^2}\right)\wtPsi_{\tp}(\tE).\label{eq.41}
\ee
The Lagrange density of this equation can be constructed by inverting the Euler-Lagrange equation and is given by
\be
\widetilde{\mathcal{L}} = \frac{1}{2} \left(\left(\partial_\tE \wtPsi\right)^2 - \left(\partial_\tp \wtPsi\right)^2 +\frac{1}{\kappa^2}\wtPsi^2\right).\label{eq.76}
\ee
Due to the chosen parameterization of $\wtPsi_\tp(\tE)$, its uniqueness as solution of (\ref{eq.41}) is satisfied when the initial data $(\wtPsi,\partial_\tp\wtPsi)$ are selected on a hypersurface of constant $\tp$. So far the field configuration is based only on the equation (\ref{accelerated_diff.-equation}), thus the shape of the fields $\wtPsi_\tp(\tE)$ is supposed not to change under translation in (momentum) space. Therefore, from the homogeneity of the (momentum) space, Noether's theorem provides us with a conserved current given by 
\be
\tC_{ij} = \frac{\partial \widetilde{\mathcal{L}}}{\partial (\partial_i \wtPsi)}\partial_j \wtPsi - \eta_{ij} \widetilde{\mathcal{L}},\quad\text{with}\, \{i,j\}=\{\tE,\tp\}.
\ee
Above, we adopt the metric signature $(+,-)$. 
Particularly in our case, for the Lagrangian density (\ref{eq.76}), the quantities that are ``momentum'' independent are
\ba
\tC_{\tp\tp} &=& \frac{1}{2} \left(\left(\partial_\tE \wtPsi\right)^2 + \left( \partial_\tp \wtPsi\right)^2 +\frac{1}{\kappa^2}\wtPsi^2\right) \\
\tC_{\tp\tE} &=& -\left(\partial_\tE \wtPsi\right)\left( \partial_\tp \wtPsi\right) \label{Theta_pE}
\ea

Translating this to the Hamiltonian description is a straightforward procedure. The conjugate momentum for $\wtPsi$ is defined as
\be
\wtPi(\tE) =\partial_\tp \wtPsi_\tp(\tE),
\ee
thus leading to the Hamiltonian
\be
X =\frac{1}{2} \int d\tE \left((\partial_\tE \wtPsi)^2 + \wtPi^2 + \frac{1}{\kappa^2}\wtPsi^2 \right),
\ee
in agreement with $\tC_{\tp\tp}$.

After we have defined the creation and annihilation operators, we can then express the Hamiltonian in terms of those operators
\be
X= \int d\ttt\,  \tx_t\left(\Phi_{x}^\dagger(\ttt)\,\, \Phi_{x}(\ttt) +\frac{1}{2}\left[\Phi_{x}(\ttt),\Phi_{x}^\dagger(\ttt)\right]\right) \label{Ham-pos}
\ee
The second term is proportional to $\delta(0)$, an infinite c-number. It is the sum over all modes of the zero-point ``positions'' $x_t/2$. We cannot avoid the existence of this term, since our treatment resembles that of harmonic oscillator and the infinite c-number term is the field analogue of the harmonic oscillator zero-point energy. 
In case of real Klein-Gordon fields, which we investigate here, this term is identically zero. However, it should be addressed the general case in which this term survives.

By similar logic applied to the Hamiltonian $X$, we can construct an operator corresponding to the one showed in Eq. (\ref{Theta_pE})
\be
T = - \int d\tE\,\, \wtPi \partial_\tE \wtPsi = \int d\ttt \,  \ttt\,\Phi_{x}^\dagger(\ttt)\,\, \Phi_{x}(\ttt)  \label{T-tim}
\ee 
Unlike the position case, no $\delta(0)$ term appears here. The modes $\ttt$ and $-\ttt$ compensate each other, so the vacuum is translationally invariant.

\end{document}